\documentstyle[12pt]{article}

\author{OSCAR JOFRE$^{1,2}$ and CARMEN N\'U\~NEZ$^{1,2,3}$\\
{}~\\
{}~\\
$^1$ Instituto de Astronom\'\i a y F\'\i sica del Espacio.\\
C.C. 67 - Suc. 28, (1428) Buenos Aires.\\
$^2$ Departamento de F\'\i sica, Facultad de Ciencias \\
Exactas y Naturales, Universidad de Buenos Aires, Argentina.\\
and\\
$^3$ International Centre for Theoretical Physics\\
P.O.Box 586 - 34100 Trieste, Italy}
\title{An Exact Cosmological Solution to String Theory
}
\begin{document}

\maketitle
\begin{abstract}
A homogeneous anisotropic four dimensional spacetime with Lo- rentzian
signature is constructed from an ungauged WZW model based on a
non-semisimple Lie group. The associated non-linear $\sigma $-model
describes string propagation in an expanding-contracting universe with
antisymmetric tensor and dilaton backgrounds. The current algebra of SL(2,R)$%
\times $R is constructed in terms of two free boson fields and two
generalized parafermions, or four free bosons with background charge. This
representation is used to study the string spectrum in the cosmological
background.\newpage\
\end{abstract}

\section{Introduction}

String propagation on a four dimensional plane wave background was recently
described with a WZW model constructed on the centrally extended Euclidean
group in two dimensions \cite{nappi}. The corresponding current algebra was
considered by Kiritsis and Kounnas \cite{parecidos1}, who used the
representations to construct the spectrum and scattering of strings moving
in gravitational wave backgrounds.

In this article, we present an example of a homogeneous, anisotropic,
four-dimensional space-time, with Lorentz signature, that can be constructed
from an ungauged WZW model based on a non-semisimple Lie group. This
space-time, with the corresponding antisymmetric tensor and dilaton fields,
can be considered as an exact cosmological solution to string theory. The
associated $\sigma $-model describes string propagation on an expanding and
contracting space-time which begins from a collapsed state, (zero volume),
and recollapses after a period of time proportional to the level $k$ of the
affine Kac-Moody algebra. This geometry is not asymptotically flat, and thus
a well defined scattering matrix cannot be constructed.

We study the current algebra corresponding to $SL(2,R)\times R$, which can
be reduced to the central extension of the 2-d Euclidean group $E_2$ ,
through an unconventional contraction.The conformal field theory (CFT)
description of the model reveals several differences from current algebras
previously considered in the literature\cite{sfetsos}-\cite{olive}. A
systematic description of current algebras based on non semisimple groups
was performed in \cite{nureidin}.

The algebra may be described in terms of two free bosonic fields without
background charge, plus two generalized parafermionic fields \cite{lykken},
or equivalently in terms
of four free bosons with background charge. We construct one representation
which may be used to study the spectrum and scattering amplitudes of bosonic
strings moving in the cosmological background.

The central charge of the CFT turns out to be non-integer (in general), and
depends on the level of the affine algebra. This differs from what happens
to other similar constructions considered so far, in which the central
charge is integer and equals the dimension of the group manifold \cite
{sfetsos}-\cite{olive}. The reason being that the bilinear form entering the
operator product expansion of the current algebra is different from the
metric on the Lie algebra, i.e., the bilinear form which raises and lowers
group indices \cite{nureidin}.

The paper is organized as follows. In Section 2 we construct the WZW model
on the group $SL(2,R)\times R$ and identify the background fields from the
associated $\sigma $ model. In Section 3 we examine some issues of duality
in the $\sigma $ model picture. The Sugawara construction on the
nonsemisimple group is performed in Section 4 where the central charge is
computed. Finally, in Section 5 we analize further the structure of the
current algebra mapping the currents into bosonic and parafermionic fields.
We also construct irreducible representations of the $SL(2,R)\times R$ Lie
algebra.

\section{The WZW model on SL(2,R)$\times $R}

Let us consider the WZW model constructed on a certain non-semisimple Lie
algebra of dimension four. The $SL(2,R)$ generators $J,P_{1,}P_2$ , satisfy
the algebra,

$$
\left[ P_a,P_b\right] =-\Lambda \epsilon _{ab}J\ \qquad \left[ J,P_a\right]
=\epsilon _{ab}P_b
$$
for $\Lambda \neq 0$, and due to the well-known ambiguity of the
two-dimensional angular momentum \cite{jackiv}, $J$ may be replaced by $J-sT$
, and $\Lambda $ by $\Lambda /s$, with $T$ a central extension. When $s$ is
set to infinity, the central extension of the two-dimensional Euclidean
group $E_2^c$ is obtained \cite{nappi}. However, instead of taking that
contraction, we redefine $P_a$ as $\sqrt{\Lambda }P_a$ , and use this
algebra, ($SL(2,R)\times R$), to construct the WZW model. Namely,

\begin{equation}
\label{algebra}\left[ P_a,P_b\right] =-\epsilon _{ab}J\ \qquad \left[
J,P_a\right] =\epsilon _{ab}P_b\qquad \left[ T,J\right] =\left[ T,P_a\right]
=0
\end{equation}

In general, given a Lie algebra with generators $T_a$ (here $%
T_a=P_1,P_2,J,T^{}$) and structure constants \ $f_{ab}^c$, to define a WZW
model one needs a bilinear form $\Omega _{ab}$ in the generators $T_a$ ,
which is symmetric, invariant

\begin{equation}
\label{inva}f_{ab}^d\Omega _{cd}\ +f_{ac}^d\Omega _{bd}=0
\end{equation}
and non-degenerate, so that there exists an inverse matrix $\Omega ^{ab},$
to raise and lower group indices. Therefore,

\begin{equation}
\label{wzw}S_{WZW}(g)=\frac 1{4\pi }\int_\Sigma d^2\sigma \Omega
_{ab}A_\alpha ^aA^{b\alpha }+\frac i{12\pi }\int_Bd^3\sigma \epsilon
_{\alpha \beta \gamma }A^{a\alpha }A^{b\beta }A^{c\gamma }\Omega
_{cd}f_{ab}^d
\end{equation}
where the fields $A_\alpha ^a$ are defined through $g^{-1}\partial _\alpha
g=A_\alpha ^aT_a$ . Here $B$ is a three-manifold with boundary $\partial
B=\Sigma $ , and $g$ is a map of $\Sigma $ to the Lie group, extended to a
map from $B$ . In order to construct the WZW action, a necessary condition
is the existence of the invariant metric $\Omega _{ab}$. Usually for
semisimple groups one can take the Cartan-Killing form $\tilde \Omega
_{ab}=f_{ac}^df_{bd}^c,$which is equivalent to $TrT_aT_b$ , with the trace
taken in the adjoint representation \cite{goddard}. However, for
non-semisimple groups this quadratic form is degenerate,
\begin{equation}
\label{metrsing}\tilde \Omega _{ab}=2\left(
\begin{array}{cccc}
1 & 0 & 0 & 0 \\
0 & 1 & 0 & 0 \\
0 & 0 & -1 & 0 \\
0 & 0 & 0 & 0
\end{array}
\right)
\end{equation}
(the only non-zero structure constants are $f_{23}^1=-f_{13}^2=-f_{12}^3=1$%
). Nevertheless, the Lie algebra (\ref{algebra}) has a non-degenerate
invariant metric, namely, the most general solution of equation (\ref{inva}%
),
\begin{equation}
\label{metrreg}\Omega _{ab}=k\left(
\begin{array}{cccc}
1 & 0 & 0 & 0 \\
0 & 1 & 0 & 0 \\
0 & 0 & -1 & 0 \\
0 & 0 & 0 & \lambda
\end{array}
\right)
\end{equation}
where $k$ and $\lambda $ are free parameters. The metric on the Lie algebra
has Lorentzian signature if $\lambda >0$, and that will be the signature of
the space-time described by the corresponding $\sigma $ model.

In order to write the WZW action we need to construct the group elements by
exponentiating the algebra and parametrizing the group manifold with
coordinates $t,x,y$ and $z.$ I.e., by writing the elements of the group as
\begin{equation}
\label{g}g=e^{xP_1}e^{tJ}e^{yP_1+zT}
\end{equation}
and using the relations%
$$
e^{-tJ}P_1e^{tJ}=\cos t\cdot P_1-\sin t\cdot P_2
$$
$$
e^{-yP_1}P_2e^{yP_1}=\cosh y\cdot P_2+\sinh y\cdot J
$$
$$
e^{-yP_1}Je^{yP_1}=\cosh y\cdot J+\sinh y\cdot P_2
$$
together with%
$$
\partial _\alpha e^H=\int_0^1dxe^{xH}\partial _\alpha He^{(1-x)H}
$$
We can then compute

\begin{equation}
\label{gdg}
\begin{array}{c}
g^{-1}\partial _\alpha g=(\partial _\alpha y+\cos t\partial _\alpha
x)P_1+(\sinh y\partial _\alpha t-\sin t\cosh y\partial _\alpha x)P_2+ \\
\\
+(\cosh y\partial _\alpha t-\sin t\sinh y\partial _\alpha x)J+\partial
_\alpha zT
\end{array}
\end{equation}
from which we can read off the elements of the algebra $g^{-1}\partial
_\alpha g=A_\alpha ^aT_a$ ,%
$$
A_\alpha ^1=\partial _\alpha y+\cos t\partial _\alpha x
$$
$$
A_\alpha ^2=\sinh y\partial _\alpha t-\sin t\cosh y\partial _\alpha x
$$
$$
A_\alpha ^3=\cosh y\partial _\alpha t-\sin t\sinh y\partial _\alpha x
$$
$$
A_\alpha ^4=\partial _\alpha z
$$
Thus, the terms that are integrated in the action (\ref{wzw}) may be
computed to be

$$
\Omega _{ab}A_\alpha ^aA^{b\alpha }=k(A_\alpha ^1A^{1\alpha }+~A_\alpha
^2A^{2\alpha }-~A_\alpha ^3A^{3\alpha }+\lambda A_\alpha ^4A^{4\alpha })
$$
$$
{}
$$
$$
=k(-\partial _\alpha t\partial ^\alpha t+\partial _\alpha x\partial ^\alpha
x+2\cos t\partial _\alpha x\partial ^\alpha y+\partial _\alpha y\partial
^\alpha y+\lambda \partial _\alpha z\partial ^\alpha z)
$$
and%
$$
\begin{array}{c}
\epsilon _{\alpha \beta \gamma }A^{a\alpha }A^{b\beta }A^{c\gamma }\Omega
_{cd}f_{ab}^d=3k\epsilon _{\alpha \beta \gamma }A^{1\alpha }(A^{2\beta
}A^{3\gamma }-A^{3\beta }A^{2\gamma }) \\
\end{array}
$$
$$
\begin{array}{c}
=6k\epsilon ^{\alpha \beta \gamma }\sin t\partial _\alpha y\partial _\beta
t\partial _\gamma x \\
\end{array}
$$
which can be written in the form%
$$
\begin{array}{c}
=6k\epsilon ^{\alpha \beta \gamma }\partial _\alpha (\cos t\partial _\beta
y\partial _\gamma x) \\
\end{array}
$$
Therefore, the Wess-Zumino term may be reduced to an integral over $\Sigma $%
, without introducing singularities,
$$
\Gamma =6k\int_Bd^3\sigma \epsilon ^{\alpha \beta \gamma }\partial _\alpha
(\cos t\partial _\beta y\partial _\gamma x)=6k\int_\Sigma d^2\sigma \epsilon
^{\beta \gamma }\cos t\partial _\beta y\partial _\gamma x
$$
Finally the action looks like

$$
S_{WZW}=\frac k{4\pi }\int_\Sigma d^2\sigma [-\partial _\alpha t\partial
^\alpha t+\partial _\alpha x\partial ^\alpha x+\partial _\alpha y\partial
^{_\alpha }y+2\cos t\partial _\alpha x\partial ^\alpha y+\lambda \partial
_\alpha z\partial ^\alpha z+\
$$
\begin{equation}
\label{sigma}+2i\epsilon ^{\beta \gamma }\cos t\partial _\beta y\partial
_\gamma x]
\end{equation}

One may read off the space-time metric, antisymmetric tensor and dilaton
fields by identifying the WZW action with the $\sigma $ model action

$$
S=\int d^2\sigma [G_{\mu \nu }\partial _\alpha X^\mu \partial ^\alpha X^\nu
+iB_{\mu \nu }\epsilon _{\alpha \beta }\partial ^\alpha X^\mu \partial
^\beta X^\nu +\Phi ^{(2)}R]
$$
where $X^\mu =\left( t,x,y,z\right) .$ The space-time geometry is described
by a Lorentz signature metric, (for $\lambda >0$), which is homogeneous but
anisotropic,
\begin{equation}
\label{espt}G_{\mu \nu }=\left(
\begin{array}{cccc}
-1 & 0 & 0 & 0 \\
0 & 1 & \cos t & 0 \\
0 & \cos t & 1 & 0 \\
0 & 0 & 0 & \lambda
\end{array}
\right)
\end{equation}
The only non-zero component of the antisymmetric field is $B_{xy}=\cos t,$
and there is also a constant dilaton background field due to the homogeneity
of the group manifold. This metric defines a cosmological model of type I,
according to the Bianchi classification, with $t$ playing the role of time
parameter, running in the range $0<t<\pi .$ At $t=0,$ the universe begins in
a collapsed state, since the determinant of the metric vanishes. At $t=\pi ,$
it recollapses again, because of the same reason. A factor $k/4\pi $ has
been suppressed from the metric (\ref{espt}), and so the actual time scale
for the expansion and recontraction is proportional to $k$. The maximum
spatial volume of this universe is reached at $t=\pi /2$ and is again of
order $k$ and proportional to $\lambda ,$ (which may be thought of as a
scale factor in the $z$ direction). The non vanishing components of the
Riemann tensor are $R_{1212}=R_{1313}=-\frac 14$ and $R_{1213}=R_{2323}=%
\left( \frac{\sin t}2\right) ^2.$ The Ricci tensor is $R_{\mu \nu }=-\frac
12\left( G_{\mu \nu }-\lambda \delta _\mu ^z\delta _\nu ^z\right) $ and the
scalar curvature $R=-\frac 32$ is constant.

This model, being a WZW model, is conformally invariant, and thus the
background satisfies the $\beta -$function equations of the non-linear $%
\sigma -$model to all orders in the $\alpha ^{\prime }$ expansion, (here $%
\alpha ^{\prime }=1/k)$. Thus, the solution differs from the homogeneous
Bianchi geometries recently found in references \cite{veneziano} as
solutions of the first order $\beta -$functions. The central charge receives
quantum corrections, therefore, unlike the models considered so far \cite
{sfetsos}-\cite{olive}, it does not equal the dimension of the group
manifold and depends on the level $k.\,$

Exact metric and dilaton backgrounds were found in reference \cite{sfetsos2}%
, using the conformal invariance of the $SL(2,R)\times
SO(1,1)^{(d-2)}/SO(1,1)$ coset models. However, the metod introduced by
Sfetsos is incapable of determining the exact antisymmetric tensor.

Consistent string propagation requires unitarity at the quantum level in
addition to conformal invariance of the corresponding non linear $\sigma$
model \cite{balog}. However, before examining the conditions for decoupling
of zero norm states, we analyze below some issues of duality in the $\sigma $
model picture.

\section{Some Duality Considerations.}

The action (\ref{sigma}) has several Killing symmetries, i.e., there are
three explicit isometries realized by translations $x\rightarrow x+a,$ $%
y\rightarrow y+b,$ $z\rightarrow z+c,$ with $a,b,c$ constants. So, there is
in principle an $O(3,3)$ duality symmetry. Let us analyze a duality
transformation in an arbitrary direction, namely in the plane defined by $x$
and $y.$ We first make a rotation $(x,y)\rightarrow (x^{\prime },y^{\prime
}),$ as%
$$
x^{\prime }=\cos \rho \cdot x-\sin \rho \cdot y
$$
$$
y^{\prime }=\sin \rho \cdot x+\cos \rho \cdot y
$$
with $\rho $ an arbitrary angle in the range $-\pi /2\leq \rho \leq \pi /2.$
Now we can make the duality transformation in the $x^{\prime }$ direction,
characterized by $\rho .$ The dual metric, antisymmetric tensor and dilaton
fields will depend on this free parameter $\rho ,$ and they are given by%
$$
\tilde G_{xx}=\frac 1{G_{xx}}=\frac 1{1+\sin 2\rho \cos t}
$$
\begin{equation}
\label{gdual}\tilde G_{xy}=\frac{B_{xy}}{G_{xx}}=\frac{\cos t}{1+\sin 2\rho
\cos t}
\end{equation}
$$
\tilde G_{yy}=G_{yy}-\frac{G_{xy}^2-B_{xy}^2}{G_{xx}}=\frac 1{1+\sin 2\rho
\cos t}
$$
\begin{equation}
\label{bdual}\tilde B_{xy}=\frac{G_{xy}}{G_{xx}}=\frac{\cos 2\rho \cos t}{%
1+\sin 2\rho \cos t}
\end{equation}
\begin{equation}
\label{dilatondual}\tilde \Phi =\Phi -\ln \left( G_{xx}\right) =\Phi -\ln
[1+\sin (2\rho )\cos t]
\end{equation}
where $G_{ij},B_{ij},$ and $\Phi $ are expressed in the coordinates $%
x^{\prime }$ and $y^{\prime }.$ The determinant and the scalar curvature of
the dual metric are%
$$
\det \tilde G=\frac{-\lambda (\sin t)^2}{(1+\sin 2\rho \cos t)^2}
$$
\begin{equation}
\label{rdual}\tilde R=\frac{1-7\cos (4\rho )+8\sin (2\rho )\cos t}{4\left(
1+\sin (2\rho )\cos t\right) ^2}
\end{equation}
and they show that the spacetime begins at $t=0$ from a collapsed state,
(zero volume). When the duality is performed in the direction determined by $%
\rho =\pi /4,$ there is no initial curvature singularity since $\tilde
R=(\sec (\frac t2))^2.$ When $t\rightarrow \pi ,$ $\det \tilde G$ and $%
\tilde R\rightarrow \infty $. For $\rho =-\pi /4$ the determinant diverges
when $t\rightarrow 0$ and vanishes for $t\rightarrow \pi ,$ (the spacetime
recollapses), while the curvature $\tilde R$ diverges for $t=0$ but it is
finite when $t=\pi .$ Recall that the original spacetime has no curvature
singularities.

 From eqs. (\ref{gdual}), (\ref{bdual}) and (\ref{dilatondual}), the
background may be seen to be self-dual for the particular values $\rho
=0,\pm \pi /2,$ i.e., when the duality is performed in the original $x$ or $%
y $ directions. Obviously, the same behaviour occurs if the duality is
performed in the $z$ direction, for $\lambda =1.$

\section{Current Algebra of the Conformal Model}

Current algebra is a useful tool to understand conformal field theories and
string theory \cite{kac}. The WZW models are simple because they realize
current algebra as its full symmetry. The action (\ref{wzw}) is invariant
under an infinitesimal transformation of the form

$$
g\longrightarrow g+\overline{\epsilon }g+g\epsilon
$$
where $\epsilon $$(z)=\epsilon ^a(z)T_a$ and $\overline{\epsilon }(\bar z)=
\overline{\epsilon }^a(\bar z)T_a,$ in complex coordinates $(z,\bar z).$ The
Noether currents associated to this symmetry, and to some Lie algebra
element, are

$$
J_z^a=\frac 1{4\pi }\Omega ^{ab}A_{zb}\qquad ,\qquad J_{\bar z}^a=\frac
1{4\pi }\Omega ^{bc}A_{\bar zb}V_c^a
$$
so that $J(z)=J_z^aT_a$ and $\bar J(\bar z)=J_{\bar z}^aT_a.$ $V_c^a$ is
defined through
$$
V_b^aT^b=g^{-1}T^ag
$$
$J_z^a$ and $J_{\bar z}^a$ are holomorphic and antiholomorphic currents,
respectively. These bosonic currents satisfy two copies of the current
algebra given by the following operator product expansion (OPE) \cite{witt},

\begin{equation}
\label{JJ}J_a(z)J_b(w)=\frac{\Omega _{ab}}{\left( z-w\right) ^2}+f_{ab}^c
\frac{J_c(w)}{\left( z-w\right) }+regular
\end{equation}
where $J_a=(P_1,P_2,J,T)$. The bilinear form $\Omega _{ab}$ must be
symmetric and invariant and the Jacobi identity states that $%
f_{abc}=f_{ab}^d\Omega _{cd}$ is completely antisymmetric. In our case $%
\Omega _{ab}$ is given by (\ref{metrreg}). (In the semisimple cases $\tilde
\Omega _{ab}$ would be used instead of $\Omega _{ab}$ ).

Once we have the current algebra, we can construct the stress tensor that is
bilinear in the currents,

$$
T(z)=L^{ab}:J_aJ_b:(z)
$$
with $L^{ab}$ a symmetric matrix determined by requiring that $T\left(
z\right) $ realizes the Virasoro algebra, i.e.,

\begin{equation}
\label{TT}T(z)T(w)=\frac{c/2}{\left( z-w\right) ^4}+\frac{2T(w)}{\left(
z-x\right) ^2}+\frac{\partial T(w)}{\left( z-w\right) }+regular
\end{equation}
and that the currents $J^a(z)$ are primary fields of conformal weight $1$
with respect to the stress tensor $T(z),$ i.e.,

\begin{equation}
\label{TJ}T(z)J^a(w)=\frac{J^a(w)}{\left( z-w\right) ^2}+\frac{\partial
J^a\left( w\right) }{\left( z-w\right) }+regular
\end{equation}
Equation (\ref{TJ}) implies that the current symmetry remains unchanged in
the quantum theory. Therefore, we get the following equations for the matrix
$L^{ab},$%
$$
L^{cb}f_{ba}^e+L^{eb}f_{ba}^c=0
$$
$$
2L^{cb}\Omega _{ba}+L^{bd}f_{ab}^ef_{ed}^c=\delta _a^c
$$
The first equation is equivalent to (\ref{inva}). Thus, $L^{ab}$ has the
same form as $\Omega ^{ab}.$ The second equation leads, uniquely, to

\begin{equation}
\label{ELE2}L_{ab}=2(k+1)\left(
\begin{array}{cccc}
1 & 0 & 0 & 0 \\
0 & 1 & 0 & 0 \\
0 & 0 & -1 & 0 \\
0 & 0 & 0 & \frac{k\lambda }{k+1}
\end{array}
\right)
\end{equation}
and $L^{ab}$ is the inverse of $L_{ab},$%
\begin{equation}
\label{ele3}L^{ab}=\frac 12\Omega ^{ab}-\delta L^{ab}\qquad with\qquad
\;\delta L^{ab}=\frac 1{2(k+1)}\left(
\begin{array}{cccc}
1 & 0 & 0 & 0 \\
0 & 1 & 0 & 0 \\
0 & 0 & -1 & 0 \\
0 & 0 & 0 & 0
\end{array}
\right)
\end{equation}
It may be seen from the last equation (\ref{ele3}), that the stress tensor
receives quantum corrections to its classical value $\frac 12\Omega ^{ab}.$
Then the central charge $c$ will receive quantum corrections as well, namely

$$
c(L)=2\Omega _{ab}L^{ab}=2\Omega _{ab}(\frac 12\Omega ^{ab}-\delta
L^{ab})=4-2\Omega _{ab}\delta L^{ab}
$$
$$
c=4-\frac 3{k+1}
$$

Recall that for non compact groups, in particular for the present $SL(2,R)$
case, the replacement $k\rightarrow -k$ has to be performed \cite{dixon};
then

\begin{equation}
\label{carga}c=4+\frac 3{k-1}
\end{equation}

Therefore, the stress tensor may be written as

\begin{equation}
\label{T}T(z)=\frac{-1}{2(k-1)}\left[ :P_1^2:+:P_2^2:-:J^2:\right] -\frac
1{2k\lambda }:T^2:
\end{equation}

\section{Representations of the Current Algebra}

We will proceed to analyze the structure of $SL(2,R)\times R$ further, by
bosonizing the Cartan subalgebra generated by $J$ and $T.$ We can define $%
J^{+}$ and $J^{-}$ as a linear combination of $J$ and $T,$%
\begin{equation}
\label{a1}J^{+}=C_1J+C_2T
\end{equation}
$$
J^{-}=C_3J+C_4T
$$
where the $C_i$ are coefficients to be determined by requiring that $J^{+}$
and $J^{-}$ diagonalize the Cartan subalgebra, i.e.,
$$
J^{+}\left( z\right) J^{+}\left( w\right) =\frac{-1}{\left( z-w\right) ^2}%
+regular
$$
\begin{equation}
\label{cartan}J^{-}\left( z\right) J^{-}\left( w\right) =\frac \mu {\left(
z-w\right) ^2}+regular
\end{equation}
$$
J^{+}\left( z\right) J^{-}\left( w\right) =regular
$$
where $\mu =+1$ for Lorentzian and $-1$ for Euclidean signature, as we shall
see below. From the OPEs above we get three equations relating the
coefficients $C_i,$%
$$
C_1^2-\lambda C_2^2=-\frac 1k
$$
\begin{equation}
\label{cs}C_3^2-\lambda C_4^2=\frac \mu k
\end{equation}
$$
C_1C_3-\lambda C_2C_4=0
$$
Thus, we may express $J^{+}$ and $J^{-}$ in terms of two bosonic fields $x^0$
and $x^3,$ as

\begin{equation}
\label{a2}J^{+}\left( z\right) =\partial x^3~,~J^{-}\left( z\right)
=\partial x^0
\end{equation}
In order to reproduce the OPE's (\ref{cartan}), the two bosons must have
propagators

$$
\langle x^3\left( z\right) x^3\left( w\right) \rangle =-\mu ~\langle
x^0\left( z\right) x^0\left( w\right) \rangle =-\ln \left( z-w\right)
{}~~~,~~\langle x^0\left( z\right) x^3\left( w\right) \rangle =0
$$

We define, in addition, two operators that will act as raising and lowering
generators%
$$
P^{\pm }=P_1\pm iP_2
$$
Using the current algebra given by the OPE (\ref{JJ}), we may calculate
\begin{equation}
\label{jp1}J^{+}\left( z\right) P^{\pm }\left( w\right) =\mp iC_1\frac{%
P^{\pm }\left( w\right) }{\left( z-w\right) }+regular
\end{equation}
\begin{equation}
\label{jp2}J^{-}\left( z\right) P^{\pm }\left( w\right) =\mp iC_3\frac{%
P^{\pm }\left( w\right) }{\left( z-w\right) }+regular
\end{equation}
so that $P^{\pm }$ are charged under the Cartan subalgebra. Similarly, we
compute
\begin{equation}
\label{pp1}P^{+}\left( z\right) P^{-}\left( w\right) =\frac{-2k}{\left(
z-w\right) ^2}+\frac{2iJ\left( w\right) }{\left( z-w\right) }+regular
\end{equation}
\begin{equation}
\label{pp2}P^{\pm }\left( z\right) P^{\pm }\left( w\right) =regular
\end{equation}
Now, $P^{\pm }\left( z\right) $ can be represented in terms of $x^0$ and $x^3
$ as
\begin{equation}
\label{j}P^{\pm }\left( z\right) =:e^{\pm i\left( C_1x^3-C_3x^0\right)
}:V^{\pm }\left( z\right)
\end{equation}
Then, the OPE's (\ref{jp1}) and (\ref{jp2}) imply that $V^{\pm }$ do not
depend on $x^0$ and $x^3.$ Defining $X^{-}=C_1x^3-C_3x^0,$ we find%
$$
P^{+}\left( z\right) P^{-}\left( w\right) =V^{+}\left( z\right) V^{-}\left(
w\right) .\left( z-w\right) ^{-C_1^2+\mu C_3^2}\times
$$

$$
\times \left[ 1+i\partial _wX^{-}\left( z-w\right) +\frac i2\partial
_w^2X^{-}\left( z-w\right) ^2-\frac 12\left( \partial _wX^{-}\right)
^2\left( z-w\right) ^2+...\right]
$$
and using equations (\ref{cs}) we get $C_1^2-\mu C_3^2=-1/k.$ This result
and the OPE's (\ref{pp1}) and (\ref{pp2}) imply that $V^{+}V^{-}$ must be of
the following form
\begin{equation}
\label{vv}V^{+}\left( z\right) V^{-}\left( w\right) =\left( z-w\right)
^{-\frac 1k}\left[ \frac{-2k}{\left( z-w\right) ^2}-2kAT\left( w\right)
+O\left( z-w\right) \right]
\end{equation}
where $A$ is a coefficient and $T\left( w\right) $ does not depend on $x^0$
and $x^3,$ but will contribute to the stress tensor, as we shall show below.

There is a representation of the algebra (\ref{vv}) in terms of the
generalized parafermions $\psi _K\,$ introduced by Lykken \cite{lykken},
(see also \cite{dixon}-\cite{bakas2}).These parafermions form an infinite
family of fields for non compact groups, which satisfy the following OPE

\begin{equation}
\label{ff}\psi _l\left( z\right) \psi _l^{\dagger }\left( w\right) \sim
\left( z-w\right) ^{-2\Delta _l}\left[ 1+\frac{2\Delta _l}{c_p}\left(
z-w\right) ^2T_p\left( w\right) +O\left( z-w\right) ^3\right]
\end{equation}
where $\psi _l^{\dagger }=\psi _{K-l}.$ The operator $T_p\left( z\right) $
is the stress tensor of the parafermionic model with central charge $%
c_p=2\left( K+1\right) /\left( K-2\right) .$ Then we have,
$$
T_p\left( z\right) \psi _l\left( w\right) =\frac{\Delta _l\psi _l\left(
w\right) }{\left( z-w\right) ^2}+\frac{\partial _w\psi _l\left( w\right) }{%
\left( z-w\right) }+O\left( 1\right)
$$
where $\Delta _l$ is the conformal dimension given by
\begin{equation}
\label{delta}\Delta _l=\frac{l\left( K+l\right) }K
\end{equation}
Comparing expressions (\ref{vv}) and (\ref{ff}), we find $K=2k,$ $l=1$ and $%
T\left( w\right) =T_p\left( w\right) .$ Then, $V^{\pm }$ can be represented
in terms of $\psi _1$ and $\psi _1^{\dagger },$ which, from now on, we
denote as $\psi _{\pm 1},$%
$$
V^{\pm }\left( z\right) =i\sqrt{2k}\psi _{\pm 1}\left( z\right)
$$
Finally, we are able to represent the $P^{\pm }$ currents in terms of two
bosons and two parafermions,%
$$
P^{\pm }=i\sqrt{2k}e^{\pm iX^{-}}\psi _{\pm 1}
$$

Let us express the Sugawara stress tensor (\ref{T}) in terms of these
fields. A straightforward computation shows that
\begin{equation}
\label{pp3}:P_1^2+P_2^2:=\frac 12\left( P^{+}P^{-}+P^{-}P^{+}\right) =-4k
\frac{\Delta _1}{c_p}T_p+k\left( \partial X^{-}\right) ^2
\end{equation}
and $\Delta _1/c_p=\left( k-1\right) /2k.$ The next step is to express $J^2$
and $T^2$ in terms of $\partial $$x^0$ and $\partial $$x^3$ using equations (%
\ref{a1}) and (\ref{a2}). For simplicity we can choose the coefficients $C_i$
so that terms proportional to $\partial x^0\cdot \partial x^3$ never appear
in the stress tensor. Thus, we have to choose $C_1=0,$ which implies $C_2=1/
\sqrt{k\lambda }$, $C_3=1/\sqrt{\mu k}$ and $C_4=0.$ Then,
\begin{equation}
\label{tt1}:T^2:=k\lambda :\left( \partial x^3\right) ^2:
\end{equation}
\begin{equation}
\label{jj1}:J^2:=\mu k:\left( \partial x^0\right) ^2:
\end{equation}

Finally, putting together eqs. (\ref{pp3}), (\ref{tt1}) and (\ref{jj1}) in
the expression for $T\left( z\right) ,$ eq. (\ref{T})
$$
T\left( z\right) =T_p\left( z\right) -\frac k{2\left( k-1\right) }\left(
\partial X^{-}\right) ^2+\frac{\mu k}{2\left( k-1\right) }\left( \partial
x^0\right) ^2-\frac 12\left( \partial x^3\right) ^2
$$
we observe that the parafermionic and bosonic contributions to the Sugawara
stress tensor decouple:%
$$
T\left( z\right) =T_p\left( z\right) +\frac \mu 2\left( \partial x^0\right)
^2-\frac 12\left( \partial x^3\right) ^2
$$
and the central charge of the full algebra is the sum of the central charge
of the free boson fields, which add up to $c_x=2,$ and the parafermionic
fields, which contribute $c_p=\left( 2k+1\right) /\left( k-1\right) .$ Thus,
the full central charge is given by:%
$$
c=c_x+c_p=2+\frac{2k+1}{k-1}=4+\frac 3{k-1}
$$
as in eq. (\ref{carga}), which confirms that we actually have a
representation of the original WZW model.

Once we have represented the current algebra in terms of bosonic and
parafermionic fields, we can construct irreducible representations of the \\$%
SL(2,R)\times R$ Lie algebra, that will serve as the base for the current
algebra representations.

There exist two independent Casimir operators, one is linear: $T$ , and the
other is quadratic:
$$
C^{\left( 2\right) }=\Omega _{ab}J^aJ^b=P_1^2+P_2^2-J^2+\lambda T^2=\frac
12\left( P^{+}P^{-}+P^{+}P^{-}\right) -J^2+\lambda T^2
$$
We begin by defining the eigenstates of the Cartan subalgebra%
$$
J\mid j,t\rangle =-ij\mid j,t\rangle
$$
$$
T\mid j,t\rangle =-it\mid j,t\rangle
$$
The action of $P^{\pm }$ on these states can be evaluated using the
hermiticity condition $(P^{\pm })^{\dagger }=P^{\mp }$%
\begin{equation}
\label{subebaja}P^{\pm }\mid j,t\rangle =\sqrt{C^{\left( 2\right) }-\lambda
t^2+j(j\pm 1)}\mid j\pm 1,t\rangle
\end{equation}
where $C^{\left( 2\right) }$ is the eigenvalue of the quadratic Casimir. $%
P^{\pm }$ act as raising and lowering operators, respectively.

We can distinguish various types of infinite dimensional representations
according to the existence of lowest (lw) or highest (hw) weight states. The
(hw) and (lw) representations are equivalent choices related by a discrete
symmetry: $J\rightarrow -J$. They are characterized by the values of $%
C^{\left( 2\right) },$ $t$ and $j$. Defining $c^{\left( 2\right) }\equiv
C^{\left( 2\right) }-\lambda t^2,$ we see from (\ref{subebaja}) that in
order to get (lw) or (hw) representations we must demand
\begin{equation}
\label{hwlw}c^{\left( 2\right) }+j(j\pm 1)=0
\end{equation}
for a particular value of $c^{\left( 2\right) },$ that we will assume $\leq
1/4,$ as will be clear below.

\underline{a) Lowest weight representations}.

For these representations we have $P^{+}\mid j,t\rangle =0.$ There are two
values of $j$ that satisfy this condition:%
$$
j_{\pm }^{\left( lw\right) }=\frac{-1\pm \sqrt{1-4c^{\left( 2\right) }}}2.
$$
For $j\geq j_{+}^{\left( lw\right) }$ and $j\leq j_{-}^{\left( lw\right) }$
we have $\sqrt{c^{\left( 2\right) }+j(j+1)}\in R$ , avoiding in this way the
zero norm eigenstates of the Casimir $C^{\left( 2\right) }.$ The (lw)
representations are characterized by the values
\begin{equation}
\label{jlw}j=j_{-}^{\left( lw\right) }-n
\end{equation}
for any natural number $n.$ This can be seen from the fact that acting with $%
P^{+}$ repetitively we can raise the index $j$ until the condition $%
(P^{+})^n\mid j,t\rangle =0$ is reached. The spectrum of $iJ$ is $%
j,j-1,j-2,\cdot \cdot \cdot .$

\underline{b) Highest weight representations}.

For these representations we need $P^{-}\mid j,t\rangle =0.$ In this case
there are also two values of $j$ satisfying this condition,%
$$
j_{\pm }^{\left( hw\right) }=\frac{1\pm \sqrt{1-4c^{\left( 2\right) }}}2.
$$
For $j\geq j_{+}^{\left( hw\right) }$ and $j\leq j_{-}^{\left( hw\right) }$
we have $\sqrt{c^{\left( 2\right) }+j(j-1)}\in R,$ and the (hw)
representations are characterized by the values
\begin{equation}
\label{jhw}j=j_{+}^{\left( hw\right) }+n,
\end{equation}
so that $(P^{-})^n\mid j,t\rangle =0.$ The spectrum of $iJ$ is $%
j,j+1,j+2,\cdot \cdot \cdot .$

\underline{c) Other representations.}

When the values of $j$ do not meet the values given by (\ref{jlw}) or (\ref
{jhw}) the representations are neither (lw) nor (hw) . In these cases we can
act with $P^{\pm }$ freely, and the representations are not bounded above or
below, but the square root in (\ref{subebaja}) becomes imaginary when $j$ is
in the interval $j_{+}^{\left( lw\right) }\leq j\leq j_{-}^{\left( lw\right)
}$ or $j_{+}^{\left( hw\right) }\leq j\leq j_{-}^{\left( hw\right) }.$

When $c^{\left( 2\right) }>1/4$ the square root in (\ref{subebaja}) becomes
complex, and the representations are neither (hw) nor (lw). The only
representation which has both a lowest and a highest weight is the unit-like
representation: $c^{\left( 2\right) }=j=0.$

It is useful to calculate the value of the zero mode of the Sugawara stress
tensor (\ref{T}),%
$$
L_0=\frac{-1}{2\left( k-1\right) }\left[ C^{\left( 2\right) }-\lambda
t^2+2j^2\right] +\frac 1{2k\lambda }t^2
$$

Once we have the representations of the Lie group we can use them to
construct the current algebra representations acting with the negative modes
of the currents. The vertex operators $W_i$ that create the primary states
must obey the following OPEs,
\begin{equation}
\label{vertices}J_a(z)W_i(w)=T_{ij}^a\frac{W_j(w)}{\left( z-w\right) }%
+regular
\end{equation}
where the coefficients $T_{ij}^a$ are representation matrices for the Lie
algebra. The dependence of the vertex operators on $x^0$ and $x^3$ can be
made explicit in the following way%
$$
W(z)\sim e^{-ip_0x^0+ip_3x^3}S(\psi _{\pm 1})
$$
where $S$ does not depend on the bosonic fields $x^0$ or $x^3.$ Applying the
Cartan subalgebra $J$ and $T$ (for $\mu =+1$), yields
$$
J\left( z\right) W\left( w\right) =-ip_0\sqrt{k}\frac{W\left( w\right) }{%
\left( z-w\right) }+regular
$$
$$
T\left( z\right) W\left( w\right) =-ip_3\sqrt{\lambda k}\frac{W\left(
w\right) }{\left( z-w\right) }+regular
$$
So that, in this parametrization, the identification%
$$
\begin{array}{c}
j=p_0
\sqrt{k} \\ t=p_3\sqrt{\lambda k}
\end{array}
$$
can be made. $P^{\pm }$ actually act as raising and lowering operators,%
$$
P^{\pm }\left( z\right) :e^{-ip_0x^0\left( w\right) +ip_3x^3\left( w\right)
}:\sim :e^{-i(p_0\pm \frac 1{\sqrt{k}})x^0\left( w\right) +ip_3x^3\left(
w\right) }:\psi _{\pm 1}\left( w\right) \left( z-w\right) ^{\mp p_0\frac 1{
\sqrt{k}}}+\cdot \cdot \cdot
$$
changing $j\rightarrow j\pm 1.$ In order to satisfy the OPE (\ref{vertices}%
), we must have%
$$
\psi _{\pm 1}\left( z\right) S\left( w\right) \sim \left( z-w\right) ^{-1\pm
p_0\frac 1{\sqrt{k}}}
$$
It is possible to adopt a free field realization of the current algebra (\ref
{ff}), which represents the two parafermionic currents $\psi $$_{\pm 1}$ as,
\cite{bakas22}-\cite{bakas23}%
$$
\psi _{\pm 1}=\frac 1{2\sqrt{k}}\left[ \pm \sqrt{2(k-1)}\partial _zx^1\left(
z\right) -i\sqrt{2k}\partial _zx^2\left( z\right) \right] e^{\pm i\sqrt{%
\frac 1k}x^2\left( z\right) }
$$
where $x^1$ and $x^2$ are two free bosons%
$$
\langle x^i\left( z\right) x^j\left( w\right) \rangle =-\delta ^{ij}\ln
\left( z-w\right)
$$
The parafermionic stress tensor, expressed in these bosonic fields is%
$$
T_p\left( z\right) =-\frac 12\left( \partial _zx^1\right) ^2-\frac 12\left(
\partial _zx^2\right) ^2+\frac 1{2\sqrt{k-1}}\partial _z^2x^1
$$
which is a Coulomb-gas representation with a background charge placed at
infinity. The full stress tensor is%
$$
T\left( z\right) =-\frac 12\eta _{\mu \nu }\partial _zx^\mu \partial _zx^\nu
+\frac 1{2\sqrt{k-1}}\partial _z^2x^1
$$
with $\eta _{\mu \nu }=diag\left( -,+,+,+\right) .$

Thus, the Sugawara stress tensor is written entirely in terms of four free
bosons of Lorentzian signature with a background charge. This simplifies the
treatment of physical states and the calculation of physical amplitudes,
since the well known screening operator technics \cite{dotsenko} may be
used. The above results are essential in understanding string propagation in
this cosmological spacetime background.Work in this direction is in progress.

\section{Conclusions.}

We have constructed a string theory on a homogeneous anisotropic four
dimensional spacetime from a nonsemisimple Lie group. This spacetime is an
expanding and contracting universe with constant scalar curvature.

By performing a duality transformation in an arbitrary direction in the
transverse space, we found other expanding or contracting backgrounds with
initial or final singularities. Since the duality transformations are valid
only to lowest order in the $\alpha^{\prime}$ expansion, string propagation
in these dual spaces is only consistent to this lowest order.

A Sugawara construction was performed when the Lie algebra (\ref{algebra})
possesses an invariant tensor $L^{ab}$ and the currents obey the algebra (%
\ref{JJ}) with $\Omega _{ab}$ given by equation (\ref{metrreg}). The
Virasoro central charge is given by equation (\ref{carga}). Unlike the
non-semisimple examples considered so far, the central charge does receive
1-loop corrections, thus it is non integer in general as observed in
reference \cite{figueroa}. It is possible to factorize this construction
into a standard (semisimple) Sugawara part and a nonsemisimple one (with
integral central charge). In the semi-classical limit ($k\rightarrow \infty $%
), we recover $c=4,$ and it is possible to map the theory in terms of four
free bosonic fields. In the general case, the conformal theory may be
represented by two bosons and two parafermions, or equivalently, by four
free bosons with a background charge placed at infinity. This may indicate a
connection between the conformal models corresponding to the cosmological
background and to another flat spacetime with a linear dilaton field in one
of the spatial directions. The order $K$ of the parafermion model depends on
the level of the affine Kac-Moody algebra, as well as the background charge
in the free boson representation.

\section{Acknowledgements.}

We would like to thank F. Quevedo for useful discussions and for carefully
reading the manuscript and E. Kiritsis for important remarks regarding
non-compact groups. This work was supported by CONICET and Universidad
de Buenos Aires. C.N. would like to thank Prof. A.\ Salam, IAEA and Unesco
for hospitality at ICTP.

\

\end{document}